# A PAPER-BASED MULTIPLEXED SEROLOGICAL TEST TO MONITOR IMMUNITY AGAINST SARS-COV-2 USING MACHINE LEARNING

Merve Eryilmaz[1,2], Artem Goncharov[1], Gyeo-Re Han[1], Hyou-Arm Joung[1,2], Zachary S. Ballard[1,2], Rajesh Ghosh[2], Yijie Zhang[1], Dino Di Carlo[2,3,*], Aydogan Ozcan[1,2,3,4,*]

[1]Electrical & Computer Engineering Department, [2]Bioengineering Department, [3]California NanoSystems Institute (CNSI), [4]Department of Surgery, University of California, Los Angeles, CA 90095 USA.

* ozcan@ucla.edu; dicarlo@ucla.edu

## ABSTRACT

The rapid spread of SARS-CoV-2 caused the COVID-19 pandemic and accelerated vaccine development to prevent the spread of the virus and control the disease. Given the sustained high infectivity and evolution of SARS-CoV-2, there is an ongoing interest in developing COVID-19 serology tests to monitor population-level immunity. To address this critical need, we designed a paper-based multiplexed vertical flow assay (xVFA) using five structural proteins of SARS-CoV-2, detecting IgG and IgM antibodies to monitor changes in COVID-19 immunity levels. Our platform not only tracked longitudinal immunity levels but also categorized COVID-19 immunity into three groups: protected, unprotected, and infected, based on the levels of IgG and IgM antibodies. We operated two xVFAs in parallel to detect IgG and IgM antibodies using a total of 40 µL of human serum sample in <20 min per test. After the assay, images of the paper-based sensor panel were captured using a mobile phone-based custom-designed optical reader and then processed by a neural network-based serodiagnostic algorithm. The trained serodiagnostic algorithm was blindly tested with serum samples collected before and after vaccination or infection, achieving an accuracy of 89.5%. The competitive performance of the xVFA, along with its portability, cost-effectiveness, and rapid operation, makes it a promising computational point-of-care (POC) serology test for monitoring COVID-19 immunity, aiding in timely decisions on the administration of booster vaccines and general public health policies to protect vulnerable populations.

**INTRODUCTION**

The COVID-19 pandemic, caused by the highly infectious SARS-CoV-2 virus, significantly disrupted life globally, making it one of modern history's most severe health crises [1, 2]. Despite notable progress in vaccine development and public health measures, SARS-CoV-2, alongside emerging variants, continues to pose a persistent threat to public health. [3-5] Even with widespread vaccination campaigns, the ongoing transmission of the infection can be partially attributed to populations experiencing a decline in neutralizing antibodies against SARS-CoV-2, which is directly related to the protection level of individuals [6]. This underscores the need to assess population-level immunity and better understand the current disease trends to mitigate morbidity and mortality of future pandemics. Along the same lines, the U.S. Food and Drug Administration (FDA) has also emphasized the importance of updated COVID-19 vaccines [7, 8]. As uncertainties persist, the effective monitoring of immunity levels becomes not only a scientific imperative but a critical foundation for public health strategies [9], and understanding the dynamics of population immunity would assist policymakers in making informed decisions with better insights into vaccination campaigns, booster shots, and other preventive measures.

Serological testing, a diagnostic tool for monitoring humoral immunity, involves measuring the production of various immunoglobulins (e.g., IgM, IgA, IgG) in response to a specific pathogen or antigen [10]. Commonly known as an antibody assay, serological testing is preferred in several applications, including surveillance and epidemiological assessments, selection of plasma for convalescent therapy, development of therapeutic antibodies, and assessment of the response to vaccinations or infections [11]. Studies have indicated that individuals with a fully functioning immune system develop an adaptive immune response after SARS-CoV-2 infection, triggering antiviral cellular and humoral reactions through T and B cell-mediated immunity, respectively [12] (see Figure 1a). In addition, antibodies targeting the nucleocapsid (N-protein) serve as a clear indicator of past infection, irrespective of an individual's vaccination status [10, 13-16]. The humoral response of an individual after an infection encompasses the production of antibodies targeting spike proteins (S-protein) and N-protein of SARS-CoV-2 (Figure 1b); however, COVID-19 vaccination (Figure 1c) elicits B-cells targeting only spike protein since this is the protein used in vaccine development (Figure 1d). [17, 18] Therefore, the distinction between generated antibodies due to past SARS-CoV-2 infection and those induced by COVID-19 vaccination can be achieved through serological analysis.

To date, lateral flow assays (LFAs) have been widely used for COVID-19 serology, offering rapid and simple point-of-care (POC) solutions for detecting antibodies [19-21]. LFAs typically focus on the detection of a single type of antibody (IgG or IgM) by using one type of SARS-CoV-2 protein (S-protein or N-protein), and this design limits the ability to differentiate IgG and IgM antibody levels in a given sample. This drawback is rooted in using a single type of SARS-CoV-2 protein within the LFA platform, where IgG can have affinity for both the S-protein and the N-protein; therefore, LFAs may not provide sufficient specificity to differentiate between antibodies generated from infection and those induced by vaccination. The difference between these antibody levels is essential for obtaining detailed information about the humoral immune response, especially for monitoring immunity over time. Thus, a multiplexed assay platform employing multiple SARS-CoV-2 proteins to simultaneously test IgG and IgM levels represents a promising direction to increase test specificity and provide a more comprehensive analysis of IgG and IgM levels, ultimately contributing to improved accuracy in identifying individuals with past infection vs. vaccination.

Partially motivated by this need, various COVID-19 serology tests have been developed to assess IgG and IgM levels in human blood/serum samples. These benchtop instruments, including e.g., ELISA, Luminex systems, Chemiluminescent Immunoassays (CLIA), and ELFIA [22-24] offer diverse testing options and provide quantitative results for antibody concentrations. While some of these benchtop designs could be multiplexed for IgG and IgM detection using multiple antigens, they require specialized and relatively costly benchtop equipment and reagents [25], which provide limitations for POC use and resource-limited settings. Furthermore, these instruments are not portable and have relatively long turnaround times for results (45-90 min), making them less suitable for field use.

In an effort to improve the portability, cost-effectiveness, accessibility, and accuracy of COVID-19 serology testing, here we report a paper-based multiplexed vertical flow immunoassay (xVFA), specifically designed to test IgG and IgM levels in human serum samples. The xVFA features a paper-based assay panel with multiple testing spots, offering a cost-effective and handheld platform, eliminating well-plates and benchtop instrumentation. Five structural SARS-CoV-2 proteins were employed within the xVFA platform, and the IgG and IgM signals from the xVFA were analyzed using machine learning (Figure 1e), which enabled extracting relevant features from the multiplexed xVFA sensor response to facilitate the classification of patients into three groups, "unprotected", "protected" and "infected" as categorized in Figure 2a. The first category, "unprotected", was defined by non-reactive samples for both IgG and IgM. Non-reactive IgG is a sign of unprotected status, as it is the primary antibody for long-term immunity. Moreover, non-reactivity to both IgG and IgM may indicate the absence or low abundance of antibodies against SARS-CoV-2. In contrast, "protected" individuals are expected to have some detectable levels of IgG as part of their developed long-term immunity, either after vaccination or weeks after infection [21]; therefore, we defined the second category as "protected" for serum samples that were reactive for IgG and non-reactive for IgM (see Figure 2a). The third category, "infected," was defined when both IgG and IgM were reactive. IgM antibodies are typically the first serology markers to indicate a recent or active infection, and IgG antibodies that develop soon after IgM production were considered indicative of an *ongoing or recent* infection, defining our third classification category: "infected".

In the development and blind testing of our serodiagnostic algorithm, human serum samples were chosen from a cohort of vaccinated individuals (with mRNA-COVID-19 vaccines), whose antibody levels were longitudinally monitored before and after each vaccine administration. These serum samples had undergone analysis using an FDA-approved clinical device [19], and a result exceeding 1.0 s/co was determined as reactive, whereas a result below 1.0 s/co was recorded as non-reactive for each antibody type tested separately [20] (Figure 2a). Subsequently, we pre-categorized these samples based on their IgG and IgM levels (s/co), and provided the *ground truth* labels for each serum sample in our training and testing sets. Following the establishment of these classification categories, our serodiagnostic neural network-based algorithm and xVFA platform were trained and validated using 120 xVFAs activated with 30 different serum samples collected over time from seven individuals. It was then blindly tested with 124 COVID-19 xVFA tests (i.e., 62 IgG and 62 IgM) on serum samples that were never used in the training/validation phase. Two separate xVFA cassettes were operated in parallel for IgG and IgM detection per sample (Figure 2c), and the images of the paper-based assay panels were captured using a custom-designed mobile-phone-based reader (Figures 2c-d) for subsequent analysis through a neural network-based decision algorithm, which achieved a blind testing accuracy of 89.5% in determining the immunity levels of individuals as protected, unprotected, or infected – all performed in < 20 min per test.

Our paper-based COVID-19 xVFA platform stands out as a cost-effective, portable and rapid IgG and IgM serology sensor that is multiplexed; with machine learning-based automated classification of patients, it could serve as a valuable POC tool for comprehensive immunity profiling of individuals, even in resource-poor settings, potentially providing unique and timely insights aiding in public health decision-making and intervention strategies.

## RESULTS

### Development of the paper-based xVFA for COVID-19 serological testing

The xVFA platform comprises different paper layers stacked to form a vertical fluidic network facilitating the capillary effect through the layers. The layers contain specifically designed waxed patterns that create hydrophobic and hydrophilic regions and allow for lateral spreading of the injected solution with a uniform flow across ~1x1 $cm^2$ sensing area. Such a design enables the incorporation of a 2D array of testing spots within the sensing membrane, and it achieves multiplexing with up to 100 spatially separated detection regions [26, 27]. For specific detection of the COVID-19 immune response, we functionalized the sensing membrane with 17 immunoreaction spots, including five SARS-CoV-2 proteins in duplicates (see Figure 2b), along with positive and negative control spots. The total assay takes <20 min and consists of two operation steps, including the loading of 20 μL of serum sample at the first step (a total volume of 40 μL for IgG and IgM combined) and the injection of 50 μL of AuNP conjugates for signal generation at the second step. The assay separately detects IgG and IgM levels through the parallel operation of two xVFA cassettes per patient sample. After the end of the assay, images of the activated paper-based sensing membranes for IgG and IgM panels are captured by a smartphone-based custom-designed optical reader (Figure 2d) and processed by a neural network-based serodiagnostic algorithm which infers COVID-19 immunity as one of the three pre-defined categories: (1) protected, (2) unprotected, (3) infected as illustrated in Figure 2c. The following subsections and the Methods explain the details about the selection of the target COVID-19 proteins, optimization of the deep learning-based inference algorithm, and clinical validation of the assay.

### Selection of the SARS-CoV-2 proteins as capture elements of the xVFA platform

The primary step in developing a serology assay is the selection of the antigen/protein targets that capture the targeted immunoglobulin panel. For the case of COVID-19 serology assays, structural proteins of SARS-CoV-2 were the primary proteins of interest (see Figure 2b). Since both the spike and non-spike proteins of SARS-CoV-2 are responsible for developing IgG and IgM after vaccination or infection [28], we selected five of the structural proteins: S1-subunit, S2-subunit, RBD-1, RBD-2, and N-protein for the assay panel of our xVFA. To accommodate this selection, we reserved two spots on the sensing membrane for each of the five proteins as the capture spots (n=10) and kept the remaining 7 spots for negative control spots (n=5) and positive control spots (n=2) (see Figure 2b). Negative control (PBS, pH 7.2) and positive control (goat anti-mouse IgG) spots at xVFA do not have specificity to analytes, and they were mainly used to normalize IgG and IgM signals of SARS-CoV-2 proteins, helping to mitigate inter-sensor variability.

After selecting the spots to be used, we evaluated IgG and IgM responses to the five selected SARS-CoV-2 proteins for determining the essential proteins for the xVFA assay panel. Positive control serum samples (n=3, post-vaccination) and negative control serum samples (n=3, Ig-free serum) were tested to assess proteins that selectively bind to IgM and IgG (Figure 2e). Our results revealed that all SARS-CoV-2 proteins generated the lowest IgG and IgM signals for the negative control samples, averaging 0.215± 0.085. Furthermore, it was noted that the positive control samples consistently exhibited elevated IgG and IgM signals in comparison to the signals of the negative control samples across the same COVID-19 proteins. The higher IgG signal for the positive control serum samples aligns with anticipated results since a high IgG signal serves as a reliable indicator of effective adaptive immunity following vaccination, and this alignment is particularly notable since the positive control group only included serum samples collected after the second vaccine dose. In addition to these, the lowest IgG signal from the positive control samples was observed with the nucleocapsid protein (0.316 ± 0.066), which is also expected since the control group consisted of samples from individuals vaccinated with mRNA-COVID-19 vaccines, which mainly trigger the generation of spike proteins rather than the nucleocapsid protein. Moreover, the highest IgG signal for the positive control sample group was recorded for RBD-1 (0.853 ± 0.094), followed by the IgG signals of the S2-protein, S1-protein, and RBD-2, in a decreasing order. This outcome correlates well with the SARS-CoV-2 proteins employed in commercially available devices and rapid test kits [11, 24, 29].

On the other hand, the IgM signals of the positive control samples for RBD-1 (0.464±0.07) were close to the other COVID-19 proteins (0.316 ±0.044). This close level of IgM signals across all the proteins is a reasonable observation because IgM is not the primary antibody for COVID-19 vaccinated individuals, and the abundance of IgM in serum is lower than the IgG levels observed post-vaccination [16]. Furthermore, the absence of infected individuals in the positive control serum samples could be another reason for the lower IgM signals that we observed against all the proteins.

**Training of neural network-based serodiagnostic algorithm**

Having selected the SARS-CoV-2 proteins for our paper-based xVFA panel, we next developed the neural network models to interpret multiplexed IgG and IgM responses of xVFA and correctly classify immune levels against SARS-CoV-2. Our deep learning-based serodiagnostic algorithm benefits from the inherent nonlinearities of neural networks and is trained to make comprehensive classification decisions from the collective responses of a panel of SARS-CoV-2-specific proteins, as detailed in the former section. We used machine learning for two purposes: first, to optimize the xVFA design, which included computational selection of the optimal subset of immunoreactions on the xVFA, and second, to use this optimized selection of IgG and IgM signals to automatically classify each patient sample into three pre-determined categories of COVID-19 immunity (protected, unprotected, or infected – see Figure 2a).

The neural network model used for this serodiagnostic algorithm is a shallow, fully connected network with three hidden layers. This network architecture was optimized through a 3-fold cross-validation method on a training/validation set composed of 120 xVFA tests obtained from 7 individuals as a function of time (see Figure 3a and Figures S2-S5; also refer to the Method section). The ground truth COVID-19 immunity levels/states for all the serum samples were determined as one of the three categories, unprotected ("Unp"), protected ("Pro"), or infected ("Inf"), by evaluating both IgG and IgM values measured by an FDA-approved device (Figure S1). The output layer of the neural network model consisted of three units, each corresponding to one immunity state with a sigmoid activation function. The network model's final decision on the immunity level classification was based on the output unit with the highest predicted normalized score (see the Methods section for details).

In the training phase of our serodiagnostic algorithm, we optimized the input signals to the network model through a feature selection method, termed the iterative backward feature elimination process (BFEP) [26, 27]. In each iteration of BFEP, immunoreactions were systematically excluded from the neural network input, one at a time, commencing with all the 14 IgG+IgM conditions (Figure 3d). Throughout this iterative process, the models to be compared were trained with 3-fold cross-validation using the training/validation serum set, and the most compelling feature subset, identified at the fourth iteration of the BFEP, with the highest accuracy between the predicted and ground truth immunity levels, was selected, which comprised 11 immunoreaction conditions (IgG+IgM) as summarized in Figure 3d. This optimal set of conditions included capture spots of RBD-1, RBD-2, S1-protein, S2-protein, N-protein, and negative control spots for IgG spots. Similarly, for IgM spots, the optimal model featured the spots for RBD-1, RBD-2, S1-protein, and N-protein, along with the positive control spots (see Figure 3g).

The performance of this neural network model using these 11 immunoreaction conditions (IgG+IgM) on the *validation* dataset achieved 98.3% accuracy for correctly classifying unprotected, protected, and infected serum samples (Figure 3e). Notably, the performance of this model with the optimal subset of immunoreaction conditions was higher than the model with all 14 immunoreaction conditions, which showed signs of model overfitting and classification errors in the validation set (see Figure 3f). Next, we will discuss the blind testing results of this optimized neural network model.

**Blind testing of the xVFA and the optimized serodiagnostic algorithm for monitoring COVID-19 immunity**

After validating the selection of SARS-CoV-2 proteins for our multiplexed assay panel and optimizing the neural network-based serodiagnostic algorithm to classify COVID-19 immunity, 124 COVID-19 xVFA tests on 31 additional serum samples collected at different time points from eight patients (never seen/used before) were blindly tested with the xVFA and the optimized serodiagnostic algorithm (see Figure 3b and Figures S6-S9). Each serum sample was tested twice by activating 4 xVFAs per sample, i.e., $IgG_1$, $IgM_1$, $IgG_2$, and $IgM_2$, where subscript indices refer to the first and second xVFA test. Out of the eight individuals used in the blind testing set, six had their serum samples drawn across multiple time points before and after their COVID-19 vaccination, and two others had a single serological test after exposure to SARS-CoV-2. Our xVFA, in a blinded manner, revealed the COVID-19 immunity state of these individuals with 89.5% accuracy with respect to the ground truth immunity levels established by a benchtop FDA-cleared instrument (see Figure 4g). At the cost of doubling the sample volume per test, the accuracy was further improved by using the test repeats of each sample. Due to the duplicate testing of both IgG and IgM, this approach results in 4 test combinations (i.e., $IgG_1$-$IgM_1$, $IgG_2$-$IgM_2$, $IgG_1$-$IgM_2$, $IgG_2$-$IgM_1$); these 4 combinations were treated independently by our serodiagnosis algorithm, yielding 4 predicted immunity states per sample (2 x 40 = 80 µL serum). By leveraging these 4 predictions, the immunity state of a sample can be determined as the state with the highest repetition among the 4 predictions. In the case of an even distribution between the predicted immunity states (e.g., 2 vs. 2 for unprotected and protected), the sample is labeled as "undetermined". This majority voting strategy using duplicate testing increased the accuracy of our xVFA to 90.3 %, correctly classifying COVID-19 immunity levels in 28 out of 31 serum samples (see Table S1 and Figure S10).

In our analyses to follow, we will refer to the prediction scores (PS) of the optimized neural network model as "PS-Pro," "PS-Unp," and "PS-Inf" for the protected, unprotected, and infected classes, respectively. The mean and standard deviation values of PS will be calculated using 4 repeats of each serum sample by duplicate testing of IgG and IgM panels. Our blinded test results reported in Figure 4 reveal that all six individuals before their first COVID-19 vaccination were correctly classified as unprotected (see Figures 4a-f), with a very strong and consistent PS-Unp score of 0.950 ± 0.044. Furthermore, a large number of the serum samples that were tested after receiving COVID-19 vaccination were correctly identified as protected based on their high PS-Pro scores from our optimized neural network model. These predictions also correlate well with the ground-truth CLIA results that show high IgG levels (>1 s/co) for these serum samples, indicating protection.

In addition to these, we were able to successfully track the immune response temporal dynamics for the six individuals whose blood testing was done at different time points after vaccinations. For instance, according to our xVFA inference, individual-1 (shown in Figure 4a) reached the maximum protection level against SARS-CoV-2 60 days after the vaccination, and our xVFA pointed out that this individual maintained a protected level for two months (114 days). The high PS-Pro score during this period correlated well with the ground-truth CLIA values, which confirmed that this individual was protected with high IgG levels. (Figure 4a). For the same individual-1, our serodiagnostic algorithm scored relatively higher PS-Inf, which is a rare case after receiving the first COVID-19 vaccine (day 19); this particular outcome was also in agreement with high IgM ground truth levels, which might be attributed to a high titer virus exposure [16] or other immune-related complications of this individual.

As another example, individual-2 revealed a high unprotection score (PS-Unp) of 0.998 before the vaccination; our xVFA inference showed that individual-2 reached the highest protection level after receiving the second vaccination, which was an abnormal scenario compared to other individuals who reached protection after the first dose of the COVID-19 vaccine as expected. Furthermore, the simultaneous interpretation of the dynamics of the three prediction scores (PS-Pro, PS-Unp, PS-Inf) as a function of time revealed a potential infection for individual-2. This delayed protection level for individual-2 after 50 days from the first vaccination correlated well with the ground truth serum sample measurements, which had a very high IgM level of 15.17 s/co on the second testing day, following a higher IgM level on the third testing day. After 106 days of the initial vaccine administration, individual-2 remained at a high PS-Pro level, revealing a decent protection level monitored through our xVFA platform. These time-lapse evaluations of the dynamics of the prediction scores and their correlation with the ground truth CLIA measurements confirm the proficiency of our serodiagnostic inference algorithm in detecting the infection even after COVID-19 vaccination and tracking longitudinal fluctuations in COVID-19 immunity over a long period.

Moreover, our serodiagnostic algorithm effectively classified individual-3 as protected by considering the higher value of the PS-Pro score on the particular testing day, 15 days after receiving the first COVID-19 vaccine dose, as shown in Figure 4c. A similar outcome was observed for individual-4 (Figure 4d), confirming the accuracy of our xVFA serodiagnostic algorithm, revealing high Ps-Pro scores for 109 days after the vaccination, which also aligned well with the ground-truth IgG and IgM levels at the same time points, referring a sufficient protection against SARS-CoV-2.

As another interesting observation, our xVFA serodiagnostic algorithm indicated a protected level (with a high PS-Pro) for individual-5 for a duration of 85 days; however, this person had a very low PS-Pro score of 0.1075 on day 113, indicating a rapidly declining protection. In comparison, individuals 3, 4 and 6 provided a maximum PS-Pro score of $0.998 \pm 0.001$ and a PS-Unp score of $0.007 \pm 0.005$ after a similar number of days post-vaccination, as shown in Figures 4c, 4d and 4f, respectively. The significant decrease in the COVID-19 protection level of individual-5 that we observed may be related to an interruption in the adaptive immune system against SARS-CoV-2 [30-32]; these cases, in general, would require further investigation using benchtop test equipment when conflicting inference scores between, e.g., PS-Pro and PS-Unp are observed. In fact, ground truth CLIA measurements for this individual-5 also confirmed some major immunity differences compared to the other patients. To shed more light on this, one can also quantify the differences between the protection scores (PS-Pro) of pre-vaccination testing (first testing, T1) and post-first vaccination testing (second testing, T2) by calculating $\Delta PSP = PSPro_{T2} - PSPro_{T1}$. For example, individual-5 exhibited a $\Delta PSP$ of 0.228, while individuals 4 and 6 showed $\Delta PSP$ values of 0.630 and 0.724, respectively. Although these differential $\Delta PSP$ values do not define the protection level, they could indicate a decline in adaptive immunity development as a function of time. For instance, after receiving the first dose of the COVID-19 vaccine (T2), individual-6 (with a high $\Delta PSP$ of 0.724) reached a ground truth IgG of 52.1 (s/co), whereas individual-5 (with a lower $\Delta PSP$) recorded a ground truth IgG level of 25 (s/co).

## **DISCUSSION**

Serological tests have been crucial in assessing immunity to COVID-19 by measuring protection levels against SARS-CoV-2 and identifying the infection rates of the population. The FDA granted emergency use authorizations (EUAs) for numerous test kits and laboratory devices throughout the pandemic. LFAs have become the primary technique due to their rapid application and simple readout [20, 21, 33]. However, there are limitations in monitoring COVID-19 immunity with LFAs.

Following vaccination or infection, human serum exhibits variable levels of IgG and IgM [10]. The low detection limit of LFAs may yield false-negative results for low IgG and IgM concentrations in human serum samples. For instance, our results obtained after testing the serum sample of individual-2 (50 days after the first vaccination) demonstrated the importance of multiplexed COVID-19 serology testing. Our xVFA serodiagnostic algorithm identified this individual as protected despite the presence of a ground truth IgM (2.94 s/co) level in this serum sample. If this individual were to be tested with an LFA, a false-positive result would indicate an infection for a protected individual. Regardless of the SARS-CoV-2 protein type used, any LFA, including RBD or S-1 protein, or containing only N-protein, would yield inaccurate results due to the high affinity of the proteins to both IgG and IgM. Thus, including all the structural SARS-CoV-2 proteins is critical for accurately testing COVID-19 immunity and preventing inaccurate or misleading outcomes. In addition, LFA-based COVID-19 IgG/IgM tests could not provide information about the protection level against SARS-CoV-2 or monitor COVID-19 immunity over time due to a lack of quantitative analysis and insufficient data interpretation [19].

The FDA provided a comprehensive overview of authorized serology tests for COVID-19 [34], where the predominant techniques involved chemiluminescence or fluorescence signals [22-25]. These benchtop measurement platforms offer enhanced sensitivity and specificity compared to LFAs; however, their bulky and relatively costly designs and labor-intensive operation make them unsuitable for POC settings. On the contrary, our xVFA has simplified operational steps and provides a portable and cost-effective alternative suitable for multiplexed POC testing.

One can better appreciate the advantages of the multiplexed sensing capability of our xVFA platform by directly comparing the results from the optimized neural network model against models that only use a single SARS-CoV-2 protein in response to IgG and IgM. When tested on the same blinded serum sample set, network models with single protein-based data inputs exhibit lower accuracies of ≤ 80.7% and they struggle to differentiate among protected, unprotected, or infected individuals (see Figures S11a-d and Figures S12-S17). All these inference models with single-plex data inputs failed to identify at least seven unprotected samples and made more erroneous predictions for the infected samples. These results demonstrate the importance of deep-learning-based multiplexed analysis for developing a reliable COVID-19 immunity profiling assay. We also compared our optimized neural network inference model to other machine learning models, including e.g., random forests and multivariable logistic regression (see Supplementary Information Note-1), which demonstrated inferior accuracies (≤ 83.1 %) compared to our optimized model (Figures S11e-f, Figures S12-S17).

In this work, we developed our xVFA platform using solely wild-proteins of SARS-CoV-2 on a paper-based assay panel for the following reasons: (i) inclusion of specific variant proteins of rapidly evolving SARS-CoV-2 may not be sensitive enough to monitor immune level after vaccination [35], (ii) the vaccines during Phase III clinical trials were developed with wild-type proteins of SARS-CoV-2, which remained in the formulations of up to date vaccines [36], and (iii) excluding any type of wild protein would lead to inaccuracies in monitoring of immune levels. Other studies based on multiplexed serology assays for COVID-19 immunity profiling were developed based on a limited number of SARS-CoV-2 variant proteins and a few wild proteins on the assay panel [37, 38]. However, the accuracy of their outcomes was not validated after receiving a COVID-19 vaccine or after being exposed to a different variant of SARS-CoV-2.

Furthermore, some of the recently reported multiplexed COVID-19 serology assays focused on monitoring a single sample cohort, e.g., infected or vaccinated, by detecting a single antibody-IgG [20, 37, 39]. In contrast, our work used both IgG and IgM levels of individuals from three different cohorts (protected, unprotected, and infected) and reported a deep learning-based serodiagnosis algorithm that benefitted from these IgG and IgM signals across five SARS-CoV-2 proteins, achieving competitive classification performance among different stages of COVID-19 immunity levels. The inclusion of these structural proteins of SARS-CoV-2 has been crucial in the accurate monitoring of different cases related to COVID-19. By analyzing combined signals from five different types of SARS-CoV-2 structural proteins immobilized on the xVFA sensing membrane, our serodiagnostic algorithm effectively resolved complications of COVID-19 testing, leading to clearer conclusions about the immune response of each patient.

In summary, we presented a multiplexed point-of-care vertical flow assay platform for serological testing of the human immune response to COVID-19. This platform detects both IgG and IgM levels binding to the main structural proteins of SARS-CoV-2, providing time-lapse monitoring of a patient's immune response. Multiplexed signals from xVFA were captured by a mobile phone-based, cost-effective optical reader and further processed with a neural network-based serodiagnostic algorithm that classified COVID-19 immunity (infected, protected, or unprotected) with an accuracy of 89.5%. This neural network-based COVID-19 serology assay features several advantages: (1) only 40 µL of serum sample is needed for testing; (2) readout is achieved by a compact and cost-effective smartphone-based optical reader; (3) the neural network-based algorithm rapidly performs robust classification of immunity state of the patient, including identification of infected samples and tracking of declined immunity over time; and (4) it is potentially scalable to different SARS-CoV-2 variants and updated COVID-19 vaccines. These advantages of our xVFA and its simple operation also make it an accessible sensing technology that could be used in various POC settings.

## METHODS

### Development of the paper-based x-VFA for COVID-19 serological testing

*Preparation of gold nanoparticle (AuNP) conjugates:* For IgM and IgG detection, 40 nM AuNP and 15 nM Au-NP (BBI Solutions) were used and modified with mouse anti-human IgM (Southern Biotech cat no 9022-01) and mouse anti-human IgG (Southern Biotech cat no 9042-01), respectively, to form the Au-NP-conjugates. For this reason, 1.0 mL of 40 nM AuNP was mixed with 100 µL borate buffer pH 8.5, and after adding 10 µL of anti-human IgM, the solution was incubated for 1 hour at room temperature. Similarly, for IgG conjugates, 1.0 mL of 15 nM AuNP was mixed with 100 µL borate buffer pH 8.5, and 10 µL of anti-human IgG was added. The other steps are the same to prepare both of the AuNP conjugates. Following the addition of the antibody, 11.2 µL of 10% bovine serum albumin (BSA, Thermo Scientific) was added to this solution to block the non-antibody-bonded regions on the AuNP. After 1 hour of blocking, the conjugates were washed three times initially with borate buffer pH 8.5, then tris buffer pH 7.2. then following the last washing, the final conjugate pellet was resuspended and stored in 100 µL of AuNP conjugate storage buffer (5% Trehalose, %0.5 Protein Saver, %0.2 Tween-20 and %1

TritonX-100 in borate buffer 8.5). Finally, the homogenous distribution and the concentration (O.D.) of the conjugates were analyzed using the absorbance mode of the microplate reader (Synergy H1; BioTek). Before running the assay, the concentration of Au-NP conjugate was diluted to an O.D. of 2.0 with the conjugate storage buffer.

*Selection of the SARS-CoV-2 proteins as capture elements of the xVFA serology test:* Spike protein subunit-1 (S1-protein) (Cat. No 230-30192), Spike protein subunit (S2-protein) (Cat. No 230-30163), Nucleocapsid (N-protein) (230-30194), and two distinct receptors binding domain (RBD) proteins, named as RBD-1 (Cat. No 230-30162) and RBD-2 (Cat. No 230-30193), throughout this manuscript, were purchased from RayBiotech Inc. These SARS-CoV-2 structural proteins were used as capture spots of our paper-based assay panel. Before developing the neural network-based algorithm and testing our clinical samples, we validated the specificity and accuracy of these proteins. To achieve this, we tested positive control samples (n=3) and negative control samples (n=3) and finalized the design of the sensing membrane. The positive control group samples were selected from serums of individuals vaccinated with mRNA vaccines, having both IgG and IgM titers, and negative control samples were prepared using IgG, IgM, and IgA-free serum, which was purchased from Sigma Aldrich (Cat. No S5393).

*Design of xVFA cartridge and the sensing membrane:* The xVFA cassette consists of two components, the bottom and the top cases, both manufactured using a 3D printer (FormLabs). The bottom case includes a paper-based assay panel (sensing membrane) at the top and absorption pads below the sensing membrane as fluid reservoirs. The top case is equipped with layers to distribute solutions across the entire surface area of the sensing membrane. Detailed material preparation and specification of paper layers can be found in our previous study. [27]. After selecting the SARS-CoV-2 proteins, a paper-based assay panel was prepared using 0.22 μm nitrocellulose membrane (Sartorius Cat. No 11327) and having 17 sensing spots where 1.0 mg/mL of each one of the SARS-CoV-2 proteins and 0.1 mg/mL of Goat anti-mouse IgG were dispensed (0.8 μL) as capture and positive control spots, respectively. Two different cassettes were used during the operation of the xVFA serological assay, separately targeting IgG and IgM. The combined cost of the xVFA cassettes for both IgG and IgM is < $7 per test (see Table S2).

*Assay operation:* The xVFA for COVID-19 serology testing was performed using two cassettes, one for IgG and one for IgM detection. Thereby, the following operation protocol was applied for each cassette in parallel. Before running the assay, background images of the sensing membranes were captured by inserting the bottom case on the portable smartphone-based optical reader. Once the initial imaging was completed, the assay started by attaching the first top case to the bottom case. Then, 200 μL of buffer solution (3% Tween-20, %1 Ovalbumin, 0.5% Protein saver, and 1% Trehalose in PBS pH 7.2) was added to wet and activate the sensing membrane (30 seconds). After total absorption of the buffer solution through the layers, 20 μL of the serum sample was introduced, where antibodies are bound to the SARS-CoV-2 proteins. After 1 minute of incubation, 300 μL of running buffer was added to wash excessive serum samples that did not react with the proteins on the sensing membrane (8 mins). After detaching the first top case, it was replaced with the second top case, and after adding 200 μL of running buffer solution (30 seconds), 50 μL of AuNP conjugate was added to generate the colorimetric signals. After 6 minutes of the washing step, the second top was detached, and the sensing membrane was imaged using a smartphone-based optical reader to measure the intensities of the capture spots under 525 nm light. It should be noted that 40 nM AuNP and 15 nm AuNP conjugates were used for IgM and IgG detection, respectively.

**Design of the portable smartphone-based optical reader and image processing workflow**

A smartphone-based optical reader was custom-designed, comprising a mobile phone (L.G. G7 ThinQ) that was assembled with a custom 3D-printed attachment to accommodate external optics (see Figure 2d and Table S3). These optical components included 4 LEDs (525 nm peak wavelength) arranged in a circular shape around an aperture with an external lens (Edmund Optics) in front of the mobile phone camera. For each LED, the front of the plastic enclosure was sanded to provide uniform illumination of the sensing membrane. The LEDs were soldered in series and powered by a constant current driver designed to output 20 mA current from 5V D.C. input. The reader was powered with a universal 5V power supply.

The images of the sensing membranes were captured before and after the assay activation in raw format (0.8 msec exposure time, 50 ISO) using the standard Android camera software and were transferred to a computer for further analysis. All immunoreaction spots before and after the assay activation were segmented by a custom-designed automated image segmentation code, and the assay signals were calculated from the pixel averages within circular segmentation masks as absorption values (see Figure 3c):

$$I_{j,r}^i = 1 - R_{j,r}^i = 1 - \frac{s_{j,r}^i}{b_{j,r}^i}, \; j = 1\text{-}7, \; r = 1\text{-}5, \; I \in \{IgG, IgM\},$$

where $b_{j,r}^i$ and $s_{j,r}^i$ are the assay signals before and after the assay activation, respectively, $R_{j,r}^i$ are reflection values, j is the immunoreaction type, r is the spot repeat within the immunoreaction, and i indicates the antibody target (i.e., IgG or IgM). The resulting 34 signals (i.e., 17 for IgG and 17 for IgM) were used for deep learning-based analysis to profile immune levels towards SARS-CoV-2 from the assay signals.

**Neural network-based serodiagnostic algorithm for monitoring COVID-19 immunity**

A neural network-based serodiagnostic algorithm was used to optimize the subset of immunoreaction conditions (from 14 immunoreactions) and classify patient immunity levels in response to SARS-CoV-2. The inputs of the algorithm were created by averaging similar spots within each immunoreaction type, yielding a maximum of 14 inputs (i.e., 7 for IgG and 7 for IgM). The network model was optimized through 3-fold cross-validation on a training/validation set, and the optimal model represented a shallow, fully connected neural network with three hidden layers (512, 256 and 128 units with ReLU activation function), each followed by batch normalization and 0.5 dropout rate. The output layer consisted of 3 units with sigmoid activations, which correspond to 3 different immunity classes: Protected ("Pro"), Unprotected ("Unp"), Infected ("Inf"). The model was optimized using a categorical cross-entropy loss, "Adam" optimizer, and a learning rate of 1e-4. The categorical cross-entropy loss function is defined as:

$$H(y, y') = -\frac{1}{N} \sum_{n=1}^{N} \sum_{x \in \{Pro, Unp, Inf\}} y_n(x) \log y'_n(x),$$

where $y_n$ are the binary ground truth values of the 3 immunity classes per patient sample, $y'_n$ are the model's predicted probability distributions for each of the three classes, and *N* is the batch size. The predicted final immunity class from the model was determined as the output class with the maximal probability in the (0,1) range from the sigmoid activation function.

The subset of input immunoreactions was optimized through an iterative backward feature elimination process (BFEP), and at each iteration, the model was trained through a 3-fold cross-validation. BFEP was performed on the training/validation set composed of 120 xVFA tests from 7 individuals. After the exclusion of each input feature, the model was evaluated based on the mean squared error (MSE) between the predicted and ground truth classes, and the feature whose exclusion yielded the lowest MSE was excluded from the feature set. As a result of the BFEP, the optimal subset of input immunoreactions included 11 features:

> RBD-1, RBD-2, S1-protein, S2-protein, N-protein and Negative controls for IgG; and
>
> RBD-1, RBD-2, S1-protein, N-protein, and Positive control for IgM (Figures 3d-g).

The model with this optimized subset of input features was blindly tested using patient test samples not used during the feature optimization process.

To train and test the performance of our models in classifying immunity levels against SARS-CoV-2, we used existing human serum samples already analyzed with an FDA-approved device and tested with PCR; these anonymized samples were purchased from Precision for Medicine, a biobank company. The samples were categorized into two groups. The first group consisted of serum samples from 12 individuals, whose samples were collected at varying time points after vaccination, and the second group included serum samples from 3 individuals infected with the SARS-CoV-2. We completed our clinical study by running a total of 244 xVFA tests, which included 61 serum samples separately tested for IgG and IgM in duplicates. The final dataset comprised 244 serum sample measurements, including 4 combinations between IgG and IgM signals from duplicate testing repeats of each antibody type (i.e., $IgG_1$-$IgM_1$, $IgG_2$-$IgM_2$, $IgG_1$-$IgM_2$, $IgG_2$-$IgM_1$, where subscript indices refer to the first and second xVFA test). To optimize, train, and blindly test our neural network models, we split these 244 VFA tests into training/validation and blind testing sets. The deep-learning models were optimized on the samples from the training/validation set, which was composed of 120 measurements/tests and 30 serum samples from 7 randomly selected individuals out of the 15 patients (Figure 3a and Figures S2-S5). The optimized model was trained on all samples from the training/validation set and was blindly tested on a testing set with 8 different individuals, 31 serum samples and 124 new xVFA tests (Figure 3b and Figures S6-S9). Importantly, when splitting the samples between the training/validation and testing sets, we ensured that all longitudinal immunity measurements for each individual were entirely assigned to training/validation or testing sets to avoid bias toward any specific individual.

**Supplementary Information** file includes:

- Supplementary Note 1: Random forest and logistic regression models
- Supplementary Tables 1-3
- Supplementary Figures S1-S17

# REFERENCES

1. Pollard, C.A., M.P. Morran, and A.L. Nestor-Kalinoski, *The COVID-19 pandemic: a global health crisis.* Physiol Genomics, 2020. **52**(11): p. 549-557.
2. Organization, W.H. *The impact of COVID-19 on global health goals.* 2021 [cited 2023 12/21]; Available from: https://www.who.int/news-room/spotlight/the-impact-of-covid-19-on-global-health-goals.
3. Ao, D., et al., *Strategies for the development and approval of COVID-19 vaccines and therapeutics in the post-pandemic period.* Signal Transduction and Targeted Therapy, 2023. **8**(1): p. 466.
4. Behl, A., et al., *Threat, challenges, and preparedness for future pandemics: A descriptive review of phylogenetic analysis based predictions.* Infection, Genetics and Evolution, 2022. **98**: p. 105217.
5. Ahmad F.B., C.J.A., Xu J., Anderson R.N., *COVID-19 Mortality Update.* 2022: United States. p. 493-496.
6. Bubar, K.M., et al., *SARS-CoV-2 transmission and impacts of unvaccinated-only screening in populations of mixed vaccination status.* Nature Communications, 2022. **13**(1): p. 2777.
7. FDA. *FDA Takes Action on Updated mRNA COVID-19 Vaccines to Better Protect Against Currently Circulating Variants.* 2023 [cited 2023 12/22]; Available from: https://www.fda.gov/news-events/press-announcements/fda-takes-action-updated-mrna-covid-19-vaccines-better-protect-against-currently-circulating.
8. Wiemken, T.L., et al., *Seasonal trends in COVID-19 cases, hospitalizations, and mortality in the United States and Europe.* Sci Rep, 2023. **13**(1): p. 3886.
9. Telenti, A., et al., *After the pandemic: perspectives on the future trajectory of COVID-19.* Nature, 2021. **596**(7873): p. 495-504.
10. Galipeau, Y., et al., *Humoral Responses and Serological Assays in SARS-CoV-2 Infections.* Front Immunol, 2020. **11**: p. 610688.
11. Infantino, M., et al., *The WHO International Standard for COVID-19 serological tests: towards harmonization of anti-spike assays.* Int Immunopharmacol, 2021. **100**: p. 108095.
12. Özbay Kurt, F.G., et al., *Booster dose of mRNA vaccine augments waning T cell and antibody responses against SARS-CoV-2.* Front Immunol, 2022. **13**: p. 1012526.
13. Guo, L., et al., *Profiling Early Humoral Response to Diagnose Novel Coronavirus Disease (COVID-19).* Clinical Infectious Diseases, 2020. **71**(15): p. 778-785.
14. Wang, Y., et al., *Kinetics of viral load and antibody response in relation to COVID-19 severity.* The Journal of Clinical Investigation, 2020. **130**(10): p. 5235-5244.
15. Al-Tamimi, M., et al., *Immunoglobulins response of COVID-19 patients, COVID-19 vaccine recipients, and random individuals.* PLoS One, 2023. **18**(2): p. e0281689.
16. Sette, A. and S. Crotty, *Adaptive immunity to SARS-CoV-2 and COVID-19.* Cell, 2021. **184**(4): p. 861-880.
17. Nguyen-Contant, P., et al., *S Protein-Reactive IgG and Memory B Cell Production after Human SARS-CoV-2 Infection Includes Broad Reactivity to the S2 Subunit.* mBio, 2020. **11**(5): p. 10.1128/mbio.01991-20.
18. Burbelo, P.D., et al., *Sensitivity in Detection of Antibodies to Nucleocapsid and Spike Proteins of Severe Acute Respiratory Syndrome Coronavirus 2 in Patients With Coronavirus Disease 2019.* The Journal of Infectious Diseases, 2020. **222**(2): p. 206-213.
19. Hirabidian, M., et al., *Evaluation of a rapid semiquantitative lateral flow assay for the prediction of serum neutralizing activity against SARS-CoV-2 variants.* Journal of Clinical Virology, 2022. **155**: p. 105268.
20. Kongsuphol, P., et al., *A rapid simple point-of-care assay for the detection of SARS-CoV-2 neutralizing antibodies.* Communications Medicine, 2021. **1**(1): p. 46.
21. Sisay, A., et al., *Diagnostic Performance of SARS-CoV-2 IgM/IgG Rapid Test Kits for the Detection of the Novel Coronavirus in Ethiopia.* J Multidiscip Healthc, 2021. **14**: p. 171-180.

22. Klüpfel, J., et al., *Automated detection of neutralizing SARS-CoV-2 antibodies in minutes using a competitive chemiluminescence immunoassay.* Analytical and Bioanalytical Chemistry, 2023. **415**(3): p. 391-404.
23. Ruscio, M., et al., *Analytical assessment of Beckman Coulter Access anti-SARS- CoV-2 IgG immunoassay.* Journal of Laboratory and Precision Medicine, 2020. **6**.
24. Stadlbauer, D., et al., *SARS-CoV-2 Seroconversion in Humans: A Detailed Protocol for a Serological Assay, Antigen Production, and Test Setup.* Current Protocols in Microbiology, 2020. **57**(1): p. e100.
25. Cox, A., et al., *Comparative evaluation of Luminex based assays for detection of SARS-CoV-2 antibodies in a transplantation laboratory.* J Immunol Methods, 2023. **517**: p. 113472.
26. Joung, H.-A., et al., *Point-of-Care Serodiagnostic Test for Early-Stage Lyme Disease Using a Multiplexed Paper-Based Immunoassay and Machine Learning.* ACS Nano, 2020. **14**(1): p. 229-240.
27. Goncharov, A., et al., *Deep Learning-Enabled Multiplexed Point-of-Care Sensor using a Paper-Based Fluorescence Vertical Flow Assay.* Small, 2023. **19**(51): p. 2300617.
28. Jackson, C.B., et al., *Mechanisms of SARS-CoV-2 entry into cells.* Nature Reviews Molecular Cell Biology, 2022. **23**(1): p. 3-20.
29. Ong, D.S.Y., et al., *How to interpret and use COVID-19 serology and immunology tests.* Clinical Microbiology and Infection, 2021. **27**(7): p. 981-986.
30. Liang, Z., et al., *Disturbance of Adaptive Immunity System Was Accompanied by a Decrease in Plasma Short-Chain Fatty Acid for Patients Hospitalized During SARS-CoV-2 Infection After COVID-19 Vaccination.* J Inflamm Res, 2023. **16**: p. 5261-5272.
31. Levin, E.G., et al., *Waning Immune Humoral Response to BNT162b2 Covid-19 Vaccine over 6 Months.* N Engl J Med, 2021. **385**(24): p. e84.
32. Khoury, J., et al., *COVID-19 vaccine - Long term immune decline and breakthrough infections.* Vaccine, 2021. **39**(48): p. 6984-6989.
33. Weissleder, R., et al., *COVID-19 diagnostics in context.* Science Translational Medicine, 2020. **12**(546): p. eabc1931.
34. FDA. *In Vitro Diagnostics Emergency Use Authorizations (EUAs) - Serology and Other Adaptive Immune Response Tests for SARS-CoV-2*. 2023 [cited 2023 12/25]; Available from: https://www.fda.gov/medical-devices/covid-19-emergency-use-authorizations-medical-devices/in-vitro-diagnostics-emergency-use-authorizations-euas-serology-and-other-adaptive-immune-response#isft1.
35. de Gier, B., et al., *Effects of COVID-19 vaccination and previous infection on Omicron SARS-CoV-2 infection and relation with serology.* Nature Communications, 2023. **14**(1): p. 4793.
36. Poland, G.A., I.G. Ovsyannikova, and R.B. Kennedy, *SARS-CoV-2 immunity: review and applications to phase 3 vaccine candidates.* The Lancet, 2020. **396**(10262): p. 1595-1606.
37. Beisenova, A., et al., *Machine-learning-aided multiplexed nanoplasmonic biosensor for COVID-19 population immunity profiling.* Sensors & Diagnostics, 2023. **2**(5): p. 1186-1198.
38. Stern, D., et al., *A bead-based multiplex assay covering all coronaviruses pathogenic for humans for sensitive and specific surveillance of SARS-CoV-2 humoral immunity.* Scientific Reports, 2023. **13**(1): p. 21846.
39. Cady, N.C., et al., *Multiplexed detection and quantification of human antibody response to COVID-19 infection using a plasmon enhanced biosensor platform.* Biosens Bioelectron, 2021. **171**: p. 112679.

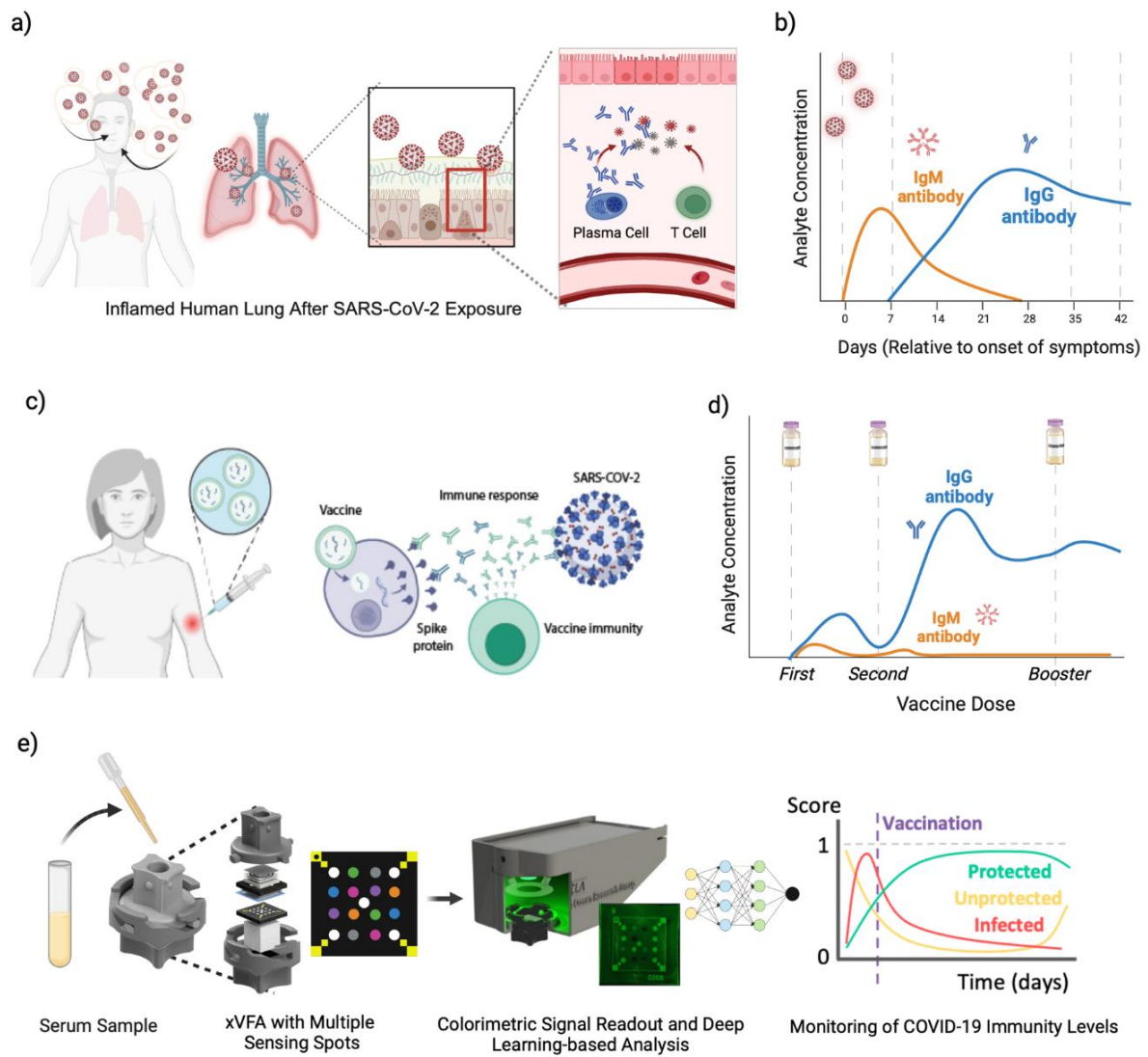

**Figure 1.** (a) Illustration of SARS-CoV-2 infection and development of COVID-19 immunity following the inflammation of the human lung (mucosal membrane). (b) IgG and IgM levels in human serum after being infected with SARS-CoV-2. (c) Illustration of humoral response after receiving a COVID-19 vaccine (m-RNA). (d) A longitudinal comparison of IgG and IgM levels in human serum after the administration of COVID-19 vaccine doses. (e) Operational steps of xVFA for COVID-19 immunity monitoring.

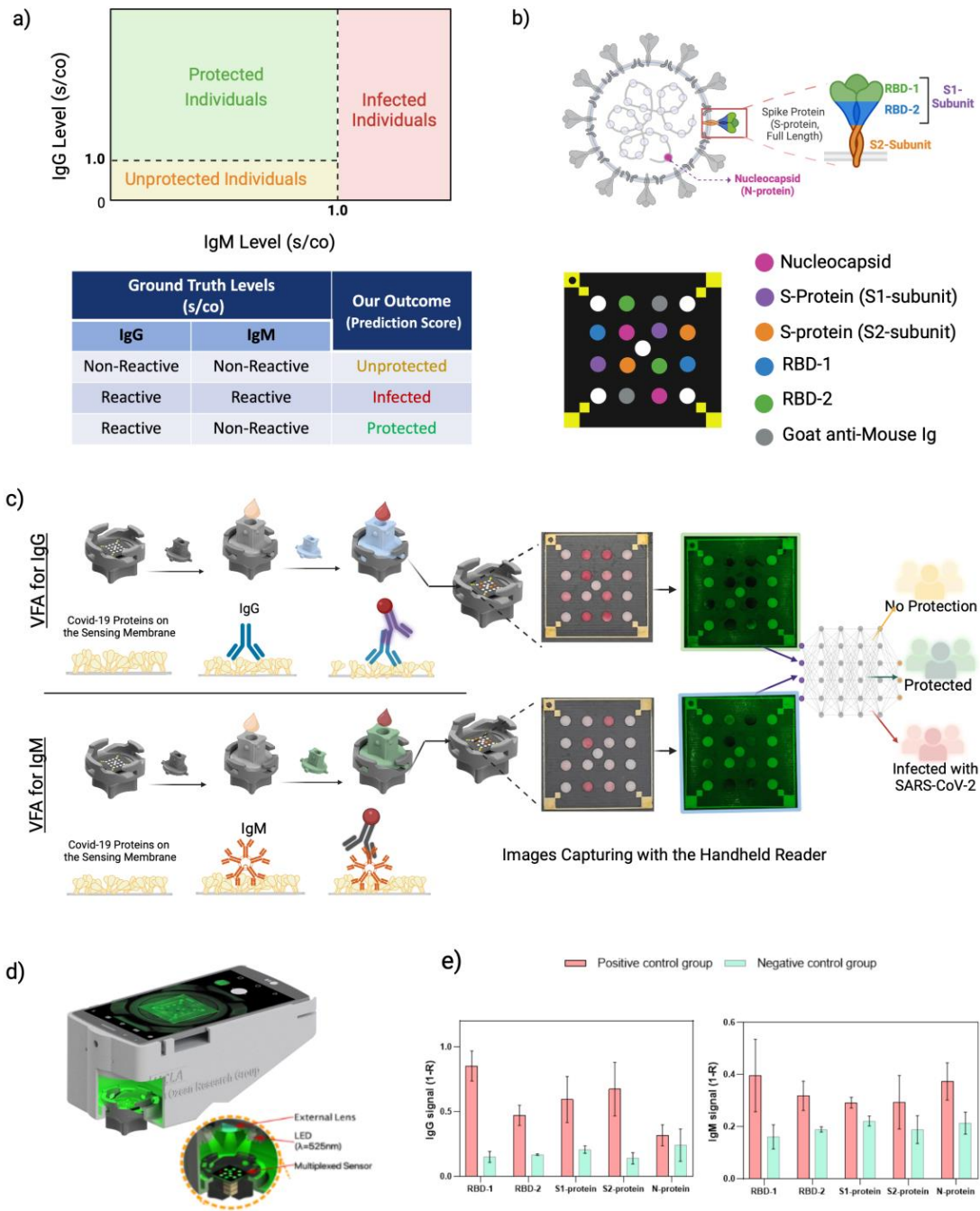

**Figure 2.** a) Categories of COVID-19 immunity levels based on the IgG and IgM levels. (b) Illustration of the SARS-CoV-2 and its structural proteins: Spike protein, Nucleocapsid, and Receptor Binding Domain (RBD); multiplexed assay design of xVFA for COVID-19 immunity monitoring. c) Steps of xVFA, running IgG and IgM antibody testing in parallel. (d) Mobile phone-based xVFA reader. (e) IgG and IgM signals of the positive control group (vaccinated individuals, n=3) and the negative control group (immunoglobulin-free serum, n=3) against five structural proteins of SARS-CoV-2.

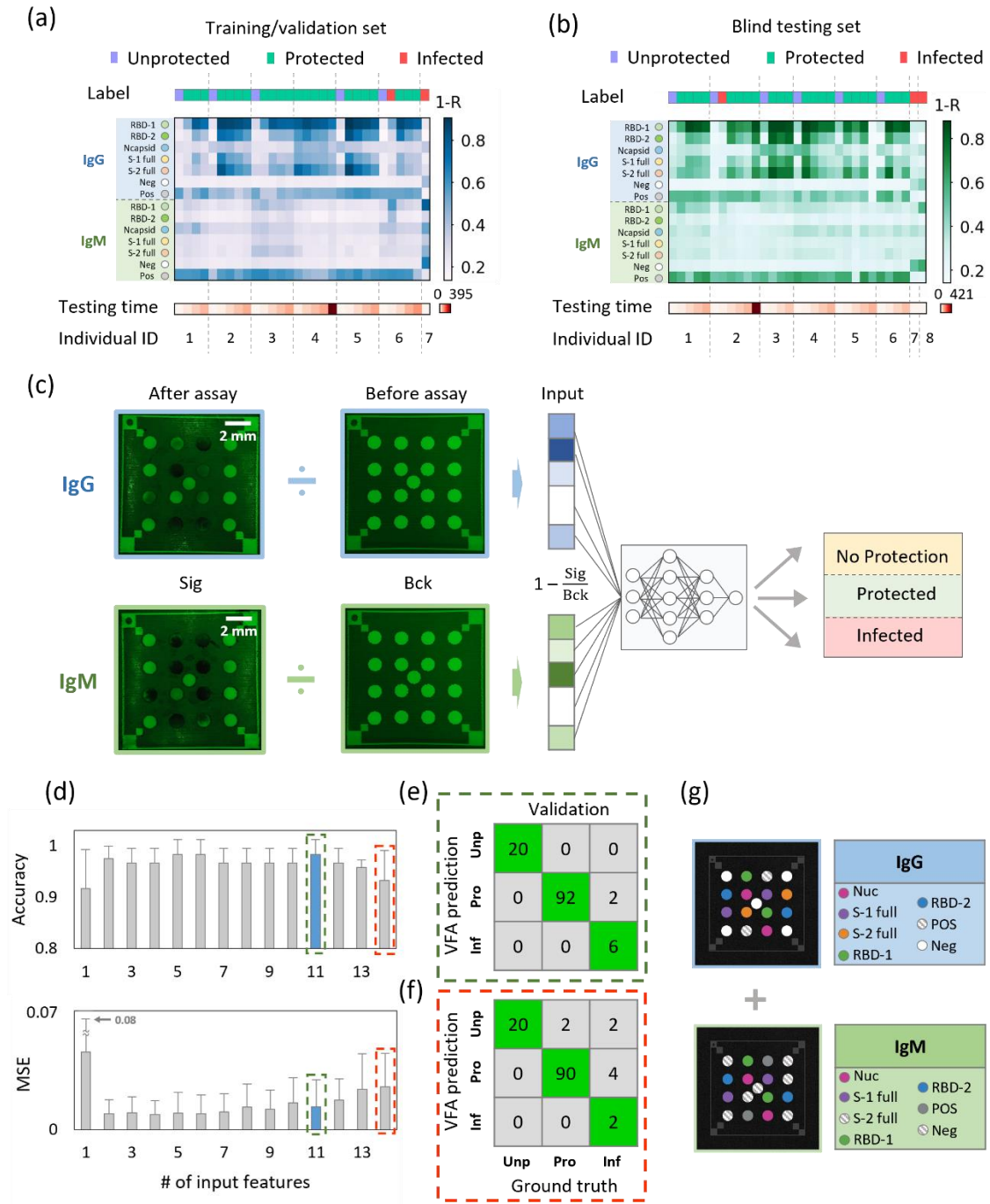

**Figure 3.** Optimization of the neural network-based xVFA testing framework. Heatmap of xVFA signals from (a) training/validation set and (b) blind testing set for the IgG and IgM panels. Signals are plotted as an average of duplicate testing. (c) Schematic of deep learning-based analysis. (d) Input feature optimization using a feed-forward feature selection process. Accuracy (top) and MSE loss (bottom) for the validation set at each iteration of the selection process. The optimal number of features was identified as 11 based on the highest accuracy and the lowest MSE loss. (e) Confusion matrix for the validation set for the optimized neural network model with the optimal subset of input features (11 features in total for IgG and IgM panels). (f) Confusion matrix for the validation set for the neural network model with all the spots used as input (14 features). (g) Optimal subsets of antigens (input features) from IgG (top) and IgM (bottom) panels.

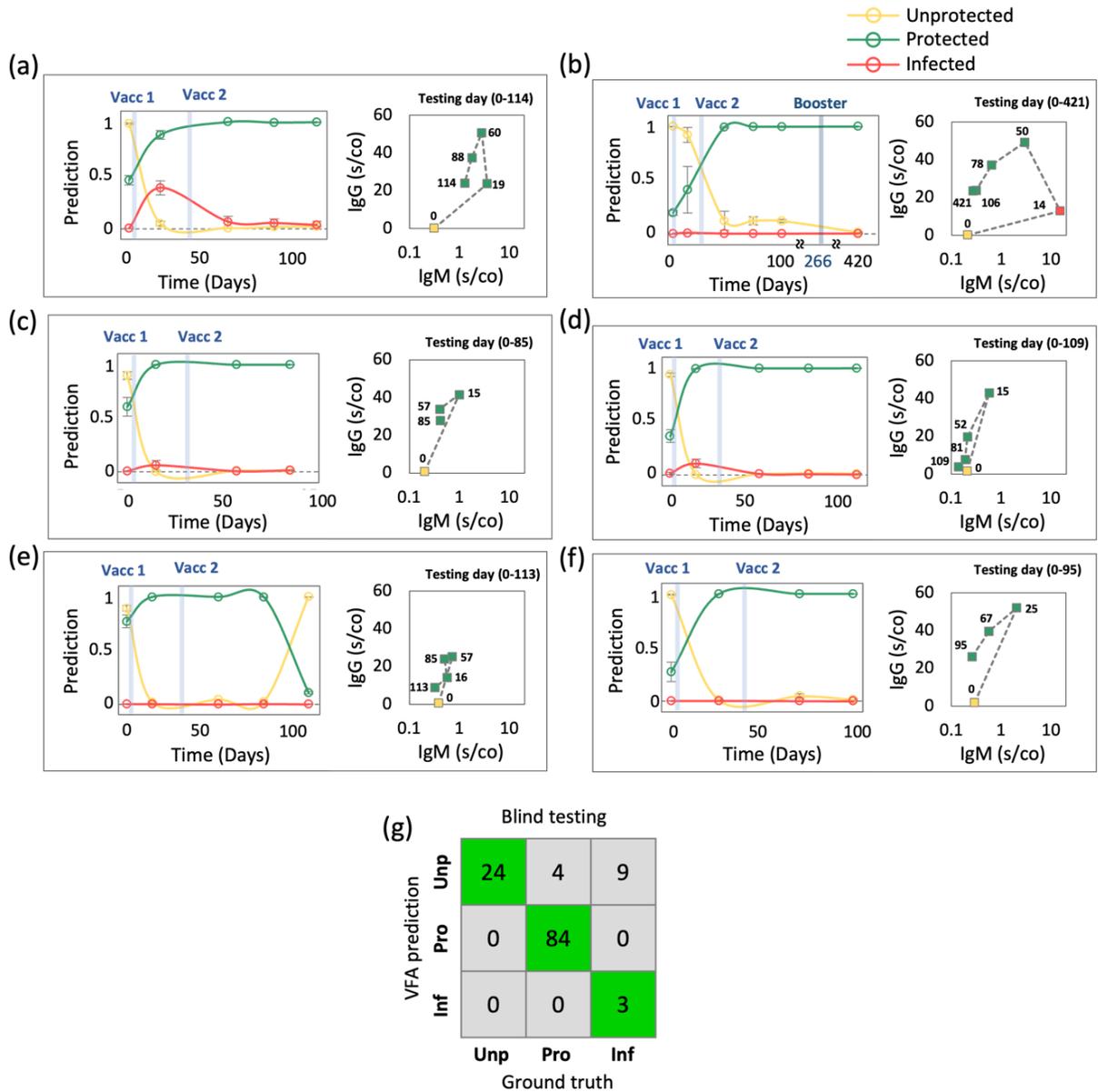

**Figure 4.** Blind testing results. (a-f) Neural network predictions of immunity level over time (left) and ground truth IgG/IgM levels (right) for 6 individuals with longitudinal sample collection. (g) Confusion matrix for all the tested serum samples from the blind testing set.